\documentclass[aps,prl,preprint]{revtex4}
\usepackage{CJK,fancyhdr}
\usepackage{graphicx}
\usepackage{booktabs}
\usepackage{amssymb,bm,mathrsfs,amscd}
\usepackage[tbtags]{amsmath}
\usepackage{lastpage}
\usepackage{color}

\newcommand{\md}{\mathrm{d}}
\newcommand{\mD}{\mathrm{D}}

\begin{document}
\title{Massless Limit of Transport Theory for Massive Fermions}
\author{Xingyu Guo}
\affiliation{Guangdong Provincial Key Laboratory of Nuclear Science, Institute of Quantum Matter, South China Normal University, Guangzhou 510006, China}

\begin{abstract}
We studied the $m=0$ limit of different components of Wigner functions for massive fermions. Comparing with the chiral kinetic theory, we separated the vanishing part and non-vanishing parts for vector and axial-vector components, up to the first order of $\hbar$. Then we discussed the possible physical meaning of the vanishing and non-vanishing parts, and their different behavior at thermal equilibrium.
\end{abstract}
\maketitle


\section{Introduction}
It is widely believed that in relativistic heavy-ion collisions(HICs), a new phase of matter called quark-gluon plasma(QGP) is created\cite{bohr_hadron_1977,venugopalan_glasma_2008}. In non-central HICs, a strong magnetic field is created by the fast moving nuclei\cite{skokov_estimate_2009,deng_event-by-event_2012}. Also, the systems have a large total angular momentum, or vorticity in the fluid picture\cite{deng_event-by-event_2017}. The effects of the magnetic field and vorticity on the QGP system are of great interest, both experimentally and theoretically, in the recent years. These include the various anomalous transport phenomena such as the chiral magnetic effect(CME)\cite{kharzeev_effects_2008,fukushima_chiral_2008,kharzeev_chiral_2016} and chiral vortical effect(CVE)\cite{son_hydrodynamics_2009,liao_anomalous_2015,kharzeev_chiral_2016}. The vorticity also has direct effect on the polarization of the hadrons produced during freeze-out\cite{liang_globally_2005,gao_global_2008,becattini_angular_2008}. The polarization of $\Lambda$ hyperons is measured by STAR collaboration\cite{the_star_collaboration_global_2017}, and the results indeed indicate a non-zero spin alignment in the direction of total angular momentum. 

Due to chiral symmetry restoration, in QGP the u and d quarks have only a very small current mass, and are often treated as chiral fermions. However, it is still worth checking whether the effect of finite mass is really neglectable. Also, the $\Lambda$ hyperons consist of s quarks, whose mass is larger and may not be ignored. Therefore the theoretic study of massive fermions is as important as that of massless ones.

Because these phenomena are closely related to the evolution of quantum systems, kinetic theory is a natural choice to study them\cite{gorbar_wigner_2017,carignano_consistent_2018,chen_calculation_2020,wei_thermal_2019}. From the Wigner function formalism\cite{wigner_quantum_1932,vasak_quantum_1987,zhuang_relativistic_1996,zhuang_relativistic_1996-1}, the kinetic theories of both massive\cite{weickgenannt_kinetic_2019,hattori_axial_2019,gao_relativistic_2019} and massless fermions\cite{gao_covariant_2017,gao_disentangling_2018,huang_complete_2018,yang_side-jump_2018,dayi_quantum_2018,kumar_wigner_2019,liu_chiral_2019,gao_dirac_2020,gao_chiral_2019} are studied. One of the advantages of Wigner function formalism is that it includes spin degree of freedom naturally, which is essential in the study of magnetic field and vorticity related phenomena. Also, it is possible to connect the kinetic description to the hydrodynamic one\cite{hidaka_nonlinear_2018,hidaka_non-equilibrium_2018,florkowski_thermodynamic_2018,kumar_wigner_2019}, as the latter one is widely used in the quantitative simulation of HIC events.

Although for spin $\frac{1}{2}$ particles, one expect that the $m=0$ limit of massive representation would connect  to massless one smoothly, it is not explicitly shown in the Wigner function formalism, especially when quantum correction is taken into consideration. This is also related to the translation or definition of different components in the theory. More detailed discussions can be found in \cite{weickgenannt_kinetic_2019,hattori_axial_2019}. 

In this paper, we propose a way of splitting the axial vector component of the Wigner function. By this splitting, we show that the massive Wigner function can indeed goes back to the massless one. We also show that there is certain physical meaning in our splitting. We will first introduce the structure of Wigner function formalism for both massive and massless fermions, and their solution up to the 1st order of $\hbar$. Then we demonstrate how to separate both the vector component and the axial vector component into a vanishing part and a non-vanishing part in $m=0$ limit, and what would be the physical meaning of the non-vanishing part. In the end we discuss briefly some possible equilibrium distributions.

\section{Covariant Wigner function} 

The covariant Wigner function for spin $\frac{1}{2}$ fermions in the presence of an external electromagnetic field is defined as\cite{vasak_quantum_1987}
\begin{eqnarray}
W_{ab}(x,p)&=& \int \md^4 y e^{i\frac{py}{\hbar}}\langle\bar\psi_b(x-{y\over 2})e^{i\frac{q}{\hbar}\int_{-{1\over 2}}^{1\over 2} dsA(x+sy)y}\psi_a(x+{y\over 2})\rangle,
\end{eqnarray}
where $q$ is the charge of fermion, and $e^{iq\int_{-{1\over 2}}^{1\over 2} dsA(x+sy)y}$ is the gauge link that ensures gauge invariance. We are considering \lq\lq{}free\rq\rq{} fermions without fermion-fermion interaction, then the Wigner function follows the kinetic equation
\begin{eqnarray}
(\gamma^\mu \Pi_\mu +\gamma^\mu\frac{i\hbar}{2}\mD_\mu-m)W=0,\label{eq:kinetic_W}
\end{eqnarray}
where
\begin{eqnarray}
\Pi_\mu&=&p_\mu -q\hbar \int_{-\frac{1}{2}}^{\frac{1}{2}}\md s s F_{\mu\nu}(x-i\hbar s\partial_p)\partial^\nu_p,\\
\mD_\mu&=&\partial_\mu-q\int_{-\frac{1}{2}}^{\frac{1}{2}}\md s F_{\mu\nu}(x-i\hbar s\partial_p)\partial^\nu_p.
\end{eqnarray}
$W$ satisfies the relationship $\gamma_0 W^\dagger \gamma_0=W$, so it can be decomposed using the Dirac matrices\cite{guo_out--equilibrium_2018}
\begin{eqnarray}
W&=&\frac{1}{4}\Big[F(x,p)+i\gamma_5 P(x,p)+\gamma_\mu V^\mu(x,p)+\gamma_\mu\gamma_5 A^\mu(x,p)+\frac{1}{2}\sigma_{\mu\nu}S^{\mu\nu}(x,p)\Big].\label{eq:decomposition}
\end{eqnarray}
Inserting eq.\ref{eq:decomposition} into eq.\ref{eq:kinetic_W}, we get the equations for all the components
\begin{eqnarray}
\Pi^\mu V_\mu&=&mF,\\
\frac{\hbar}{2} \mD^\mu A_\mu&=&mP,\\
\Pi_\mu F-\frac{1}{2}\hbar \mD^\nu S_{\nu\mu}&=&mV_\mu,\\
-\hbar \mD_\mu P+\epsilon_{\mu\nu\sigma\rho}\Pi^\nu S^{\sigma\rho}&=&2m A_\mu,\\
\frac{1}{2}\hbar(\mD_\mu V_\nu- \mD_\nu V_\mu)+\epsilon_{\mu\nu\sigma\rho}\Pi^\sigma A^\rho&=&mS_{\mu\nu},
\end{eqnarray}
and 
\begin{eqnarray}
\hbar \mD^\mu V_\mu&=&0,\\
\Pi^\mu A_\mu&=&0,\\
\frac{1}{2}\hbar D_\mu F+\Pi^\nu S_{\nu\mu}&=&0,\\
\Pi_\mu P+\frac{\hbar}{4}\epsilon_{\mu\nu\sigma\rho}\mD^\nu S^{\sigma\rho}&=&0,\\
\Pi_\mu V_\nu-\Pi_\nu V_\mu-\frac{\hbar}{2}\epsilon_{\mu\nu\sigma\rho}\mD^\sigma A^\rho&=&0.
\end{eqnarray}
These equations can be solved by expanding all the operators and functions as series of $\hbar$ and finding solutions order by order. For each order, not all the components are independent as they must follow the constraints given by the equations above. So we can use a certain number of the components to express all the others. For these \lq\lq{}free\rq\rq{} components, their corresponding kinetic equations are given by the equations of one order higher, as the derivatives always come with $\hbar$.

The 0th order solution is\cite{weickgenannt_kinetic_2019}:
\begin{eqnarray}
P^{(0)}&=&0,\\
F^{(0)}&=&m f^{(0)}_V\delta(p^2-m^2),\\
V^{(0)}_\mu&=&p_\mu f^{(0)}_V\delta(p^2-m^2),\\
S^{(0)}_{\mu\nu}&=&\frac{1}{m}\epsilon_{\mu\nu\sigma\rho}p^\sigma A^{(0)\rho}.\label{eq:S-theta}
\end{eqnarray}
Here we choose $f_V$, or $F$, together with $A_\mu$ as independent components. There is also one constraining equation for $A^{(0)}_\mu$: 
\begin{eqnarray}
p^\mu A^{(0)}_\mu=0\label{eq:constrain_A0},
\end{eqnarray}
so there are all-together 4 degrees of freedom. It is also possible to use $S_{\mu\nu}$ instead of $A_\mu$ as free components. The total degree of freedom is the same after taking into consideration all the constraint equations.

The same procedure can be taken for massless fermions. However, in massless case, the vector components $V_\mu$ and $A_\mu$ decouples from others, and at the 0th order, they can be expressed as\cite{huang_complete_2018}
\begin{eqnarray}
V^{(0)}_\mu &=&p_\mu f^{(0)}_V\delta(p^2),\label{eq:ml-0th}\\
A^{(0)}_\mu &=&p_\mu f^{(0)}_A\delta(p^2).
\end{eqnarray}

\section{Massless Limit}

At first glimpse, the $m\to 0$ limit for $V_\mu$ is quite clear, but that for $A_\mu$ is not. Actually, in the massive case, there is not a very simple and unique expression for $A_\mu$, with only one constraining equation. In order to compare with the massless case, we propose the following separation:
\begin{eqnarray}
A^{(0)}_\mu&=&(p_\mu f^{(0)}_A-\theta^{(0)}_{\mu})\delta(p^2-m^2).
\end{eqnarray}
From eq.\ref{eq:constrain_A0} we can get the relation between $f^{(0)}$ and $\theta^{(0)}_{\mu}$:
\begin{eqnarray}
(p^2f^{(0)}_A-p\cdot\theta^{(0)})\delta(p^2-m^2)=0.\label{eq:relation_A}
\end{eqnarray}
But still there is one redundant degree of freedom. For a given $A_\mu$, we can change $\theta^{(0)}_\mu$ by an arbitrary  vector that is parallel to $p_\mu$, and modify $f^{(0)}_A$ according to eq.\ref{eq:relation_A}. The new set will also give the same $A^{(0)}_\mu$. To get rid of this arbitrariness, we must fix $f^{(0)}_A$. This can be achieved by introducing an auxiliary time-like vector $n_\mu$ and requiring
\begin{eqnarray}
\theta^{(0)}\cdot n=0.
\end{eqnarray}
These will lead to the relation
\begin{eqnarray}
f^{(0)}_A&=&\frac{A^{(0)}\cdot n}{p\cdot n}.\label{eq:n-fA}
\end{eqnarray}
Different choices of $n$ corresponds to different values of $f^{(0)}_A$ and $\theta^{(0)}_\mu$. Therefore it is important to choose the $n_\mu$ with proper physical meaning. One of the nature choice is relating $n_\mu$ to the local \lq\lq{}average velocity\rq\rq{}, which corresponds to fluid velocity in ideal hydrodynamics\cite{rezzolla_relativistic_2013}. Then this can be related to the physical fact that when observing in different reference frames, the helicity of massive fermion can be different. For comparison, although we can write the same expresion for $f_V^{(0)}$ and $V^{(0)}_\mu$, $f_V^{(0)}$ actually does not depend on $n$ because $V^{(0)}_\nu$ is proportional to $p_\mu$.

In masslese case, from eq.\ref{eq:ml-0th} we can see that the relation eq.\ref{eq:n-fA} holds naturally. Now we only have to prove that $\theta^{(0)}_\mu$ vanishes when $m$ goes to zero. This can 
also be demonstrated using $n_\mu$. If we assume that the Wigner function will not diverge when $m$ goes to 0, then eq.\ref{eq:S-theta} requires that $\theta^{(0)}_\mu$ vanishes. We can rewrite it into another form to see more clearly: 
\begin{eqnarray}
\theta^{(0)}_\mu\delta(p^2-m^2)&=&-\frac{m}{2p\cdot n}\epsilon_{\mu\nu\sigma\rho}n^\nu S^{\sigma\rho}.
\end{eqnarray}
As $p\cdot n$ is generally not zero, $\theta_\mu^{(0)}$ must vanish when $m$ goes to 0.

Therefore we can see the $m\to 0$ limits for both $V^{(0)}_\mu$ and $A^{(0)}_\mu$ are the same with the massless solutions.

In the next we proceed to the 1st order components. Similar to the 0th order case, we can also use $f_V$ and $A_\mu$ as free components, then we will have\cite{gao_relativistic_2019}:

\begin{eqnarray}
P^{(1)}&=&\frac{1}{2m}\mD^\mu A^{(0)}_\mu,\\
F^{(1)}&=&m f^{(1)}\delta(p^2-m^2)-\frac{1}{2m(p^2-m^2)}\epsilon_{\mu\nu\sigma\rho}p^\mu\mD^\nu p^\sigma A^{(0)\rho},\\
V^{(1)}_\mu&=&p_\mu f^{(1)}_V\delta(p^2-m^2)+\frac{qp_\mu}{2m^2(p^2-m^2)}\epsilon_{\alpha\beta\sigma\rho}F^{\sigma\rho}p^\alpha A^{(0)\beta}\nonumber\\
&&+\frac{1}{2m^2}\epsilon_{\mu\nu\sigma\rho}\mD^\nu p^\sigma A^{(0)\rho},\\
A^{(1)}_\mu&=&(p_\mu f^{(1)}_A-\theta^{(1)}_\mu)\delta(p^2-m^2)-\frac{1}{2(p^2-m^2)}\epsilon_{\mu\nu\sigma\rho}p^\nu\mD^\sigma V^{(0)\rho},\\
S^{(1)}_{\mu\nu}&=&\frac{1}{2m}(\mD_\mu V^{(0)}_\nu-\mD_\nu V^{(0)}_\mu)+\frac{1}{m}\epsilon_{\mu\nu\sigma\rho}p^\sigma A^{(1)\rho}.
\end{eqnarray}

$V^{(1)}_\mu$ can be rewritten as
\begin{eqnarray}
V^{(1)}_\mu&=&p_\mu f^{(1)}_V\delta(p^2-m^2)-\frac{p_\mu}{2p^2 p\cdot n}\epsilon_{\alpha\beta\sigma\rho}p^\alpha n^\beta (\mD^\sigma \theta^{(0)\rho})\delta(p^2-m^2)\nonumber\\
&&+\frac{q}{2p^2}\epsilon_{\mu\nu\sigma\rho}F^{\sigma\rho}p^\nu f^{(0)}_A\delta(p^2-m^2)+\frac{1}{2 p\cdot n}\epsilon_{\mu\nu\sigma\rho}n^\nu(\mD^\rho\theta^{(0)\sigma})\delta(p^2-m^2)\nonumber\\
&&-\frac{1}{2 p\cdot n}\epsilon_{\mu\nu\sigma\rho}n^\nu p^\sigma(\mD^\rho f^{(0)}_A)\delta(p^2-m^2).\label{eq:V1}
\end{eqnarray}
In the derivation, we used the Schouten identity $p_{\mu}\epsilon_{\nu\sigma\rho\lambda}+p_{\nu}\epsilon_{\sigma\rho\lambda\mu}+p_{\sigma}\epsilon_{\rho\lambda\mu\nu}+p_{\rho}\epsilon_{\lambda\mu\nu\sigma}+p_{\lambda}\epsilon_{\mu\nu\sigma\rho}=0$ and also $\mD_\mu \delta(p^2-m^2)=\frac{2F_{\mu\nu}p^\nu}{p^2-m^2}\delta(p^2-m^2)$.
If we take the massless limit, $\theta^{(0)}_\mu$ vanishes and one can easily check that eq.\ref{eq:V1} goes back to
\begin{eqnarray}
V^{(1)}_\mu&=&p_\mu f^{(1)}_V\delta(p^2-m^2)+\frac{q}{2p^2}\epsilon_{\mu\nu\sigma\rho}F^{\sigma\rho}p^\nu f^{(0)}_A\delta(p^2-m^2)\nonumber\\
&&-\frac{1}{2 p\cdot n}\epsilon_{\mu\nu\sigma\rho}n^\nu p^\sigma(\mD^\rho f^{(0)}_A)\delta(p^2-m^2),
\end{eqnarray}
which is the same as the result for chiral fermions\cite{huang_complete_2018}. For $n_\mu$, we can use the same definition as in the 0th order case.

For $A^{(1)}_\mu$, we rewrite it as
\begin{eqnarray}
A^{(1)}_\mu&=&(p_\mu f^{(1)}_A-\tilde\theta^{(1)}_\mu)\delta(p^2-m^2)-\frac{1}{2p^2}\epsilon_{\mu\nu\sigma\rho}p^\nu\mD^\sigma V^{(0)\rho}\nonumber\\
&&-\frac{1}{2 p\cdot n}\epsilon_{\mu\nu\sigma\rho}n^\nu p^\sigma (\mD^\rho f^{(0)}_V)\delta(p^2-m^2)\label{eq:A1}\\
\tilde\theta^{(1)}_\mu\delta(p^2-m^2)&=&-\frac{m}{2p\cdot n}\epsilon_{\mu\nu\sigma\rho}n^\nu S^{(1)\sigma\rho}-\frac{m^2 q}{2p\cdot n(p^2-m^2)}\epsilon_{\mu\nu\sigma\rho}n^\nu F^{\sigma\rho}f^{(0)}_V\delta(p^2-m^2)\\
f^{(1)}_A\delta(p^2-m^2)&=&\frac{A^{(1)}\cdot n}{p\cdot n}-\frac{q}{2p\cdot n p^2}\epsilon_{\mu\nu\sigma\rho}n^\mu p^\nu F^{\sigma\rho} f^{(0)}_V\delta(p^2-m^2).
\end{eqnarray}
Again, $\tilde\theta_\mu$ should vanish in the massless limit, and the remaining part is the same as the massless expression\cite{huang_complete_2018}. Thus we have shown explicitly that up to the first order of $\hbar$, the massive Wigner function can connect to massless one continuously. These are consistent with the results found in Ref.\cite{weickgenannt_kinetic_2019,hattori_axial_2019}. Actually, at lowest order, the dipole-momentum tensor $\Sigma_{\mu\nu}$ used in\cite{weickgenannt_kinetic_2019} can be expressed as
\begin{eqnarray}
\Sigma_{\mu\nu}=\frac{1}{m^2}\epsilon_{\mu\nu\sigma\rho}p^\sigma\theta^\rho.
\end{eqnarray}
Taking the same procedure for deriving eq.\ref{eq:V1} and eq.\ref{eq:A1}, this can be shown to equal to $\frac{1}{p\cdot n}\epsilon_{\mu\nu\sigma\rho}n^\sigma A^\rho$, which is the spin tensor used in\cite{hattori_axial_2019}. When taking $m\to 0$, it will go to $-\frac{1}{p\cdot n}\epsilon_{\mu\nu\sigma\rho}p^\sigma n^\rho$, as required by\cite{weickgenannt_kinetic_2019}. Therefore the different formulations are connected, and this connection is closed related to the massless limit itself.

We should also check the meaning of $f_V$ and $f_A$. In our case
\begin{eqnarray}
f^{(0)}_V&=&\frac{V^{(0)}\cdot n}{p\cdot n},\\
f^{(0)}_A&=&\frac{A^{(0)}\cdot n}{p\cdot n},\\
f^{(1)}_V&=&\frac{V^{(1)}\cdot n}{p\cdot n}-\frac{q}{2p^2p\cdot n}\epsilon^{\mu\nu\sigma\rho}n^\mu p^\nu F^{\sigma\rho}f^{(0)}_A\delta(p^2-m^2),\\
f^{(1)}_A&=&\frac{A^{(1)}\cdot n}{p\cdot n}-\frac{q}{2p^2p\cdot n }\epsilon_{\mu\nu\sigma\rho}n^\mu p^\nu F^{\sigma\rho} f^{(0)}_V\delta(p^2-m^2).
\end{eqnarray}
These are consistent with massless case. It is already known that $f_V$ is the particle number density. In massless case, $f_A$ is the difference between left and right handed particles, or the chiral imbalance. Now that we have a continuous expression, we might call $f_A$ chiral imbalance as well, but in massive case it is not a conserved charge. However, we observe that when choosing the same $n_\mu$, the expression of $f^{(0)}_A$ become different with some literatures such as\cite{hattori_axial_2019} while they have the same massless limit. $\theta_\mu$ might be viewed as the real spin degrees of freedom. In massless case, especially in the chiral kinetic theory, it is not seen because the spins of chiral fermions are bound with their momentum. Of course, the degrees of freedom are not lost. We could discuss massless Dirac fermions, and they will have two degrees of freedom in the $S_{\mu\nu}$ component. It is only because $V_\mu$ and $A_\mu$ decouple from $S_{\mu\nu}$, $F$ and $P$ that the latter ones are not included in the usual chiral kinetic theories. On the other hand, in massive case, the chiral states are not energy eigenstates, and the spin degree of freedom is coupled to other ones, just as $S_{\mu\nu}$ is coupled to $A_{\mu}$. In fact, in massive case, it is possible to use $S_{\mu\nu}$ instead of $A_\mu$ as free components to construct the whole kinetic theory\cite{weickgenannt_kinetic_2019}.
\section{Kinetic Equations and Solutions}

The discussion above naturally leads to the question of what is the equilibrium distribution of all the components. We will only consider the 0th order components in this section. The transport equations for 0th components are:
\begin{eqnarray}
p\cdot \mD^\mu f^{(0)}_V &=&0,\\
(p\cdot\mD\theta^{(0)}_\mu-qF_{\mu\nu}\theta^{(0)\nu})-p_\mu p\cdot\mD f^{(0)}_A&=&0.
\end{eqnarray}
Without the collision term, one can not determine the true equilibrium state only from the kinetic equation. But we can still make some general discussions. The equation for $f^{(0)}_V$ is just a Boltzmann-type equation without an collision term. It would be natural to assume that in equilibrium, $f^{(0)}_V$ takes the form of usual Fermi-Dirac distribution. In the other equation, there is also a Boltzmann-type equation for $f^{(0)}_A$, coupled to a BMT equation\cite{bargmann_precession_1959} involving $\theta^{(0)}_\mu$. A straightforward guess is that $f^{(0)}_A$ also takes the Fermi-Dirac distribution, leaving the equation for $\theta^{(0)}_\mu$ as
\begin{eqnarray}
p\cdot\mD\theta^{(0)}_\mu-qF_{\mu\nu}\theta^{(0)\nu}&=&0.
\end{eqnarray}
But if $f^{(0)}_A$ really is the chiral imbalance as we supposed, with finite mass it should be dispersed over time and, at least at classical level, reaches zero at equilibrium. 

In massless situation, we have
\begin{eqnarray}
f_V^{(0)}&=&f_+^{(0)}+f_-^{(0)},\\
f_A^{(0)}&=&f_+^{(0)}-f_-^{(0)},\\
f_\chi^{(0)}&=&\frac{1}{e^{\mathrm{sgn}(p\cdot n)\frac{p\cdot n-\mu_\chi}{T}}+1},
\end{eqnarray}
where $\chi=+$ and $-$ correspond to right-handed and left-handed components, respectively. $\mu_\chi$ is the corresponding chemical potential. If the right-handed and left-handed components are balanced, we would have $\mu_+=\mu_-=\mu$, then
\begin{eqnarray}
f_V^{(0)}&=&\frac{2}{e^{\mathrm{sgn}(p\cdot n)\frac{p\cdot n-\mu}{T}}+1},\\
f_A^{(0)}&=&0.
\end{eqnarray}
This is consistent with the massive situation we just discussed.

Of course when $f^{(0)}_A=0$, there is always a trivial solution $\theta^{(0)}_\mu=0$, giving zero $A_\mu^{(0)}$. But if $A^{(0)}_\mu$ is the average spin of system, it should not always be zero. We can imagine starting from a special initial state where the spins of all particles are polarized along one direction. With a collision term, either spin itself is conserved, or it is coupled to orbital angular momentum, but the total angular momentum is conserved. Either way, it is very unlikely that the average spin will evolve to exactly zero at equilibrium. So in general we should expect nonzero $A_\mu$ even in equilibrium.
In the special case of only a constant thermal vorticity $\omega_{\mu\nu}=\frac{1}{2}(\partial_\mu \beta_\nu-\partial_\nu\beta_\mu)$, where $\beta_\mu = \frac{n_\mu}{T}$, there is a very interesting solution
\begin{eqnarray}
\theta^{(0)}_\mu&=&a\epsilon_{\mu\nu\sigma\rho}n^\nu p^\sigma \omega^{\rho\lambda}p_\lambda,\\
f^{(0)}_A&=&0,
\end{eqnarray}
where $a$ is an undetermined variable quantifying the \lq\lq{}strength\rq\rq{} of the polarization.  We can see that now $A^{(0)}_\mu$ is non-zero, but $f^{(0)}_A$ is zero. By our previous translation of $f_A$, this solution means that there is non-zero spin polarization without any chiral imbalance, which is physical.

For another special case with a homogeneous electromagnetic field and no vorticity, there is also a similar solution
\begin{eqnarray}
\theta^{(0)}_\mu&=&b F_{\mu\nu}p^\nu,\\
f^{(0)}_A&=&0.
\end{eqnarray}
By this solution we require there being no electric field $F_{\mu\nu}n^\nu=0$.

These two simplified cases show that it is at least possible to have zero chiral imbalance but non-zero average spin at equilibrium. For more complicated case with both vorticity and magnetic field, there also are solutions like\cite{liu_covariant_2020}:
\begin{eqnarray}
A_\mu&=&-\frac{1}{2m\Gamma}\epsilon_{\mu\nu\sigma\rho}p^\nu\mD^\sigma \beta^\rho-\frac{\hbar}{2(p^2-m^2)}\epsilon_{\mu\nu\sigma\rho}p^\nu\mD^\sigma V^{(0)\rho}.
\end{eqnarray}
But this solution is differernt from the above ones as now even at 0th order, $A\cdot n$ is not zero, thus giving a non-zero $f_A$.

\section{Summary}
By making a special splitting of the axial vector component of the covariant Wigner function for massive fermions, we showed that the $m=0$ limit can be taken, and the result are consistent with the massless Wigner function. We also discussed the different meaning of the splitted parts and their possible equilibrium value.

\begin{acknowledgments}
The author thanks Xinli Sheng, Yu-Chen Liu, Xu-Guang Huang and participants of the QKT2019 workshop for very helpful discussions. Part of this work is done during XG\rq{}s visit to the Institute of Theoretical Physics, Frankfurt University.

The author became aware of the related work\cite{sheng_kinetic_2020} after the completion of this study.

\end{acknowledgments}

\bibliography{quantumtransporttheory}

\end{document}